\documentclass[aip,jcp,reprint]{revtex4-1}

\usepackage{epstopdf}
\usepackage{subfigure}
\usepackage[utf8]{inputenc}
\usepackage[T1]{fontenc}
\usepackage{graphicx}
\usepackage[]{natbib}
\usepackage{pgfplots}
\usepackage[babel,german=quotes]{csquotes}
\usepackage{float}
\usepackage[]{hyperref}
\usepackage{ltablex}

\draft 

\begin{document}

\title{Intermediate state dependence of the photoelectron circular dichroism of fenchone observed via femtosecond resonance-enhanced multi-photon ionization} 

\author{Alexander Kastner}
\author{Tom Ring}
\affiliation{Institut für Physik und CINSaT, Universität Kassel, Heinrich-Plett-Strasse 40, 34132 Kassel, Germany}
\author{Bastian C. Krüger}
\author{G. Barratt Park}
\author{Tim Schäfer}
\affiliation{Institut für Physikalische Chemie, Georg-August-Universität Göttingen, Tammannstr. 6,
37077 Göttingen, Germany  }

\author{Arne Senftleben}
\author{Thomas Baumert}
\email[]{t.baumert@uni-kassel.de}
\affiliation{Institut für Physik und CINSaT, Universität Kassel, Heinrich-Plett-Strasse 40, 34132 Kassel, Germany}

\date{\today}

\begin{abstract}

The intermediate state dependence of photoelectron circular dichroism (PECD) in resonance-enhanced multi-photon ionization of fenchone in the gas phase is experimentally studied. By scanning the excitation wavelength from 359 to 431 nm we simultaneously excite up to three electronically distinct resonances. In the PECD experiment performed with a broadband femtosecond laser their respective contributions to the photoelectron spectrum can be resolved. High-resolution spectroscopy allows us to identify two of the resonances as belonging to the B- and C-bands, which involve excitation to states with 3s and 3p Rydberg character, respectively. We observe a sign change in the PECD signal depending on which electronic state is used as an intermediate. Additionally, scanning the laser wavelength reveals a decrease of PECD magnitude with increasing photoelectron energy for the 3s state.

\end{abstract}

\pacs{}

\maketitle 

\section{Introduction}

Chiral recognition in the gas phase using electromagnetic radiation is an emerging research field and promising for fundamental research as well as for applications due to the non-interacting nature of molecules in the gas phase. Progress using microwave techniques, \cite{Patterson.2013, Patterson.2013b, Shubert.2014} laser mass spectroscopy, \cite{Boesl.2013, Horsch.2012} Coulomb explosion imaging for direct absolute configuration determination \cite{Herwig.2013, Pitzer.2013, Pitzer.2016, Pitzer.2016b} and laser as well as vacuum ultra-violet synchrotron radiation based chiral recognition in the gas phase has been recently reviewed. \cite{Boesl.2016, Nahon.2015, Patterson.2014, Wollenhaupt.2016} Photoelectron angular distributions turned out to be especially sensitive and are usually measured by velocity map imaging (VMI) techniques. \cite{Chandler.1987, EppinkA.T.J.B..1997} \\ \indent
An asymmetry arising in the photoelectron angular distribution resulting from ionization of  optically active molecules with circularly polarized light was predicted by theory \cite{Ritchie.1976b} and demonstrated using single-photon ionization via synchrotron radiation. \cite{Bowering.2001}  
This circular dichroism in the angular distribution \cite{Garcia.2003} is termed photoelectron circular dichroism (PECD). \cite{Powis.2008} 
Because it arises directly from electric dipole interaction, the magnitude of PECD of up to a few ten percent typically surpasses that of other chiroptical techniques. In addition, gas phase PECD techniques can be used as highly sensitive analytic tool \cite{Lux.2012c, Lux.2015b, Lehmann.2013} with respect to investigation of enantiomeric excess, \cite{Kastner.2016, Nahon.2016} chiral mixtures, \cite{Fanood.2015} conformation of molecules 
\cite{Turchini.2013} or vibrational levels of the cation. \cite{Powis.2013, Garcia.2017} \\ \indent 
PECD is a sensitive probe of photoionization dynamics as it originates from the quantum interference of outgoing partial waves when the photoelectron is 
scattered off the chiral molecular potential. \cite{Powis.2013} Since the first demonstration 
of single-photon PECD, recently also from high harmonic generation laser sources, \cite{Ferre.2014} a variety of new observations were made and reviewed. \cite {Nahon.2015}
Theoretical modeling of PECD and experiments showed that in valence shell ionization the asymmetry will decrease for higher photoelectron energies. \cite{Powis.2000} Moreover, PECD varies with photoelectron energy including sign changes throughout the 0\textendash 10 eV kinetic energy range for single-photon ionization out of the highest occupied molecular orbital (HOMO) of fenchone. \cite{Nahon.2016} PECD was also observed when using a core-shell \textit{initial orbital} as e.g. the C 1s from the C=O group in camphor \cite{Hergenhahn.2004} and fenchone \cite{Powis.2008c} and consequently interpreted as being predominantly a final state scattering effect. \cite{Hergenhahn.2004, Lein.2014} Recently, X-ray absorption spectroscopy demonstrated site-selective excitation of the C 1s 
orbital situated at the stereocenter of fenchone. \cite{Ozga.2016} \\ \indent  
Resonance-enhanced multi-photon ionization (REMPI) gives access to \textit{electronic intermediates} and, with the help of femtosecond laser excitation and ionization, PECD has been demonstrated in a 2+1 REMPI scheme of bicyclic ketones. \cite{Lux.2012c, Lehmann.2013} As more angular momentum can be transferred in a multi-photon process in comparison to single-photon ionization, higher order nodal structures were observed. Consequently, even higher order 
nodal structures were observed in above threshold ionization \cite{Lux.2016} and experiments with longer wavelengths showed PECD also in the tunneling regime. \cite{Beaulieu.2016a} Recently, the first time-resolved experiments addressing dynamics in the intermediate have been reported. \cite{Beaulieu.2016b, Comby.2016} \\ \indent
By employing REMPI schemes, ionization via different electronic intermediates can be investigated by tuning the wavelength of the laser. The dependence of multi-photon PECD on wavelength has been investigated before. \cite{Lehmann.2013, Fanood.2016, Beaulieu.2016a} In the case of limonene, the influence of the electronic character of intermediates on PECD was interpreted to be rather unimportant. \cite{Beaulieu.2016a, Fanood.2016} Two separated contributions in the photoelectron spectrum when exciting an electron from the HOMO have been observed in the case of multi-photon ionization of camphor molecules \citep{Lehmann.2013} and assigned to excitation of distinct Rydberg states during 2+1 REMPI. \\ \indent
A continuous wavelength scan covering several excited states allows the study of the influence of the electronic character of the intermediate on PECD as well as the dependence of PECD on photoelectron energy. It can serve as a benchmark for emerging theoretical descriptions of multi-photon PECD. \cite{Goetz.2017, Lein.2014, Demekhin.2015, Lehmann.2013} \\ \indent
In this contribution, we extend previous studies on wavelength dependence with a step size smaller than the bandwidth of the femtosecond laser. We observe distinct 2+1 REMPI channels via the B- and C-band transitions, which are believed to correspond to 3s $\leftarrow$ n and 3p $\leftarrow$ n excitation, respectively. \cite{Pulm.1997} In this paper, we will refer to the upper electronic states of these transitions as 3s and 3p. The outline of the paper is as follows. In section \ref{sec:ExpSetup}, the excitation scheme as well as the two experimental setups used are described. In section \ref{sec:nsREMPI}, a fine scan of 2+1 REMPI using narrow-bandwidth
nanosecond laser excitation is shown. Results of femtosecond PECD measurements are given in section \ref{sec:fsPECD}.

\section{Experimental setup and data evaluation}
\label{sec:ExpSetup}

\begin{figure}[htb]
\subfigure[]
{\includegraphics[width=0.25\linewidth]{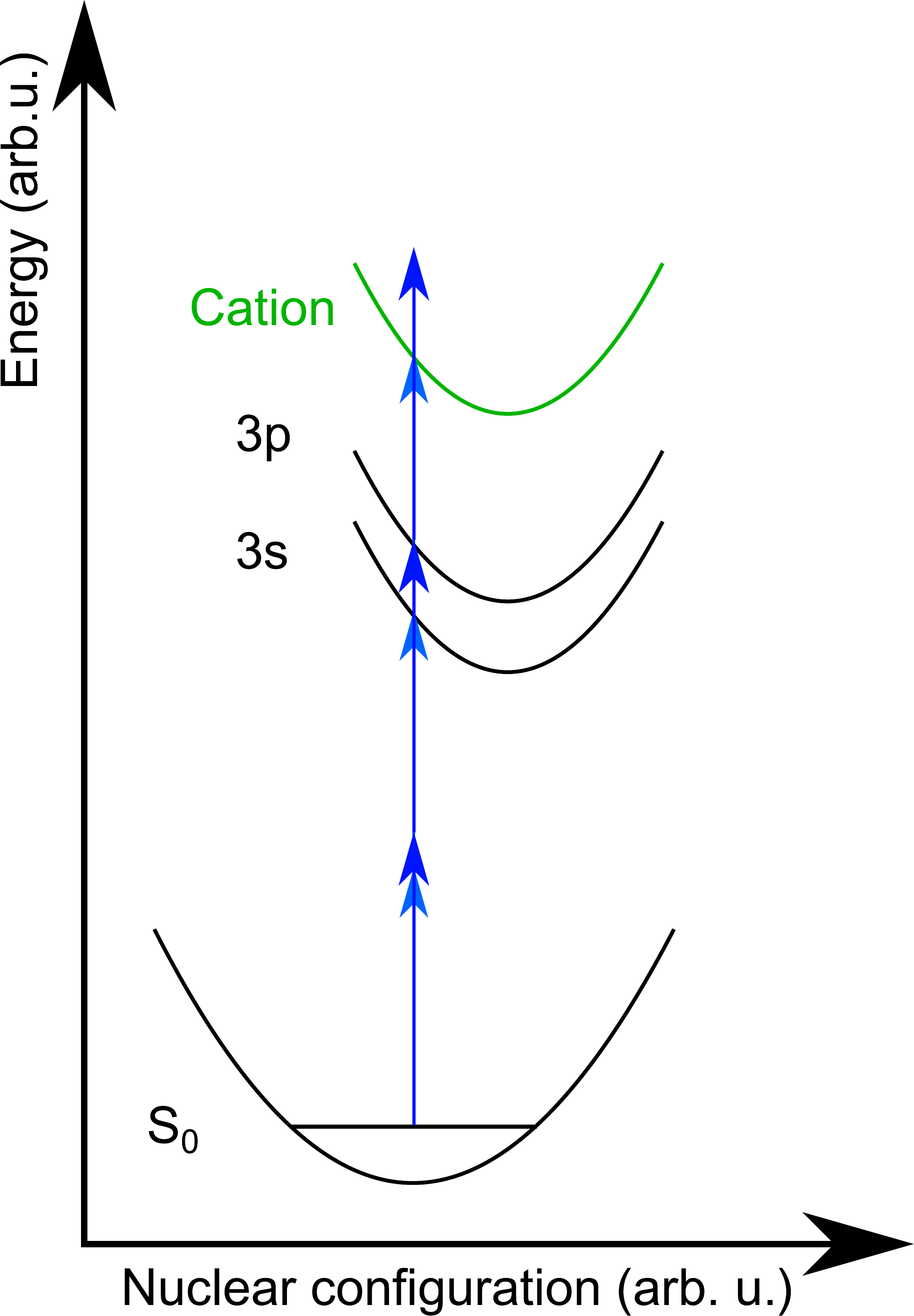}
\label{fig:ExcScheme}}
\subfigure[]
{\includegraphics[width=0.45\linewidth]{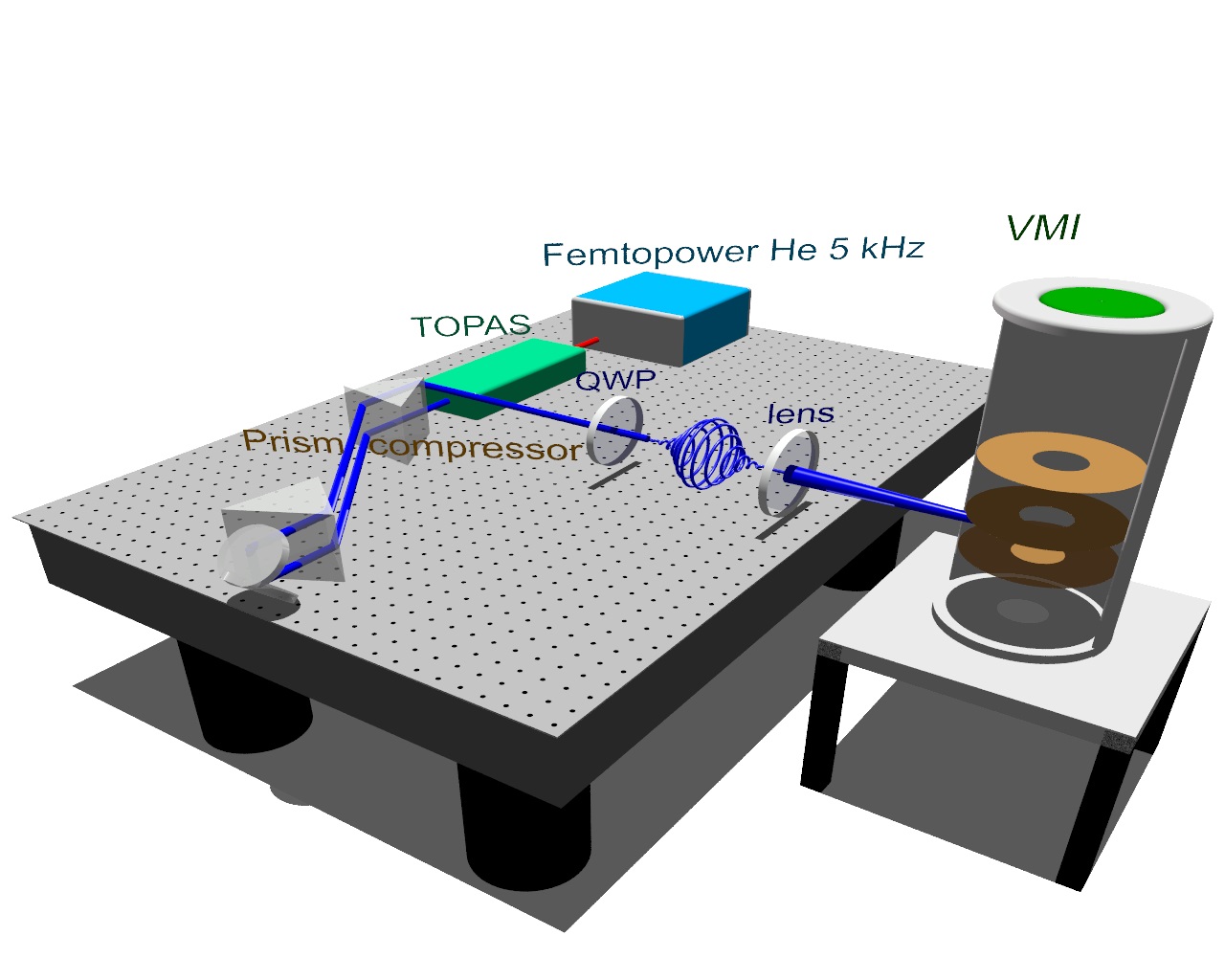}
\label{fig:ExpSetup}}
\caption{(a) Schematic excitation and ionization scheme for the 2+1 REMPI and (b) schematic layout of the femtosecond photoelectron circular dichroism experiment. A detailed description is given in the text.}
\end{figure}

The excitation and ionization scheme used for the experiments can be understood as a multi-photon ionization through excitation of intermediate states (schematically shown in figure \ref{fig:ExcScheme}). Depending on wavelength, the ionization is possible via different intermediate states promoting the photoelectrons to different kinetic energies in the continuum.

\subsection{High-resolution 2+1 REMPI experiment}
\label{sec:nsExpSetup}

The nanosecond laser setup utilized for the high-resolution 2+1 REMPI experiment in section \ref{sec:nsREMPI} is as follows. Ultra-violet (UV) pulses in the spectral region between 375 and 420 nm are provided by a frequency-doubled optical parametric oscillator (OPO) system (\textit{Continuum Sunlite Ex FX-1}) with 3 GHz bandwidth yielding pulse energies of about 4 mJ at 400 nm. The pulses are focused into the interaction region of a time-of-flight (TOF) spectrometer using an $f =$ \mbox{200 mm} lens. The wavelength of the OPO is recorded by a wavemeter (\textit{High Finesse WS7}) and the UV pulse energy is monitored by a pyroelectric detector. To resolve vibrational structure in the 2+1 REMPI, the scan shown in blue in \mbox{figure \ref{fig:Absspectra}} was acquired with a step size of  \mbox{$0.05$ cm$^{-1}$.} \\ \indent
The TOF spectrometer used in the experiment has been described earlier. \cite{Park.2016} Briefly, fenchone is evaporated by resistive heating of a reservoir coupled to a pulsed nozzle synchronized to the laser. The vapor is expanded using a backing pressure of 6 bar H$_2$. The beam is doubly skimmed and enters the ultra-high vacuum chamber, where the photoions are detected on a double stack multi-channel plate (MCP) detector with a conical anode collecting the amplified electron clouds. The integrated yield of the parent ion TOF peak is recorded using an oscilloscope (\textit{LeCroy LT344}) as a function of excitation wavelength and\textemdash due to the experimentally determined quadratic power law\textemdash \ corrected with the square of the corresponding laser power.

\subsection{Femtosecond PECD experiment}

\begin{figure}[htb]
\subfigure[]{\includegraphics[width=0.4\linewidth]{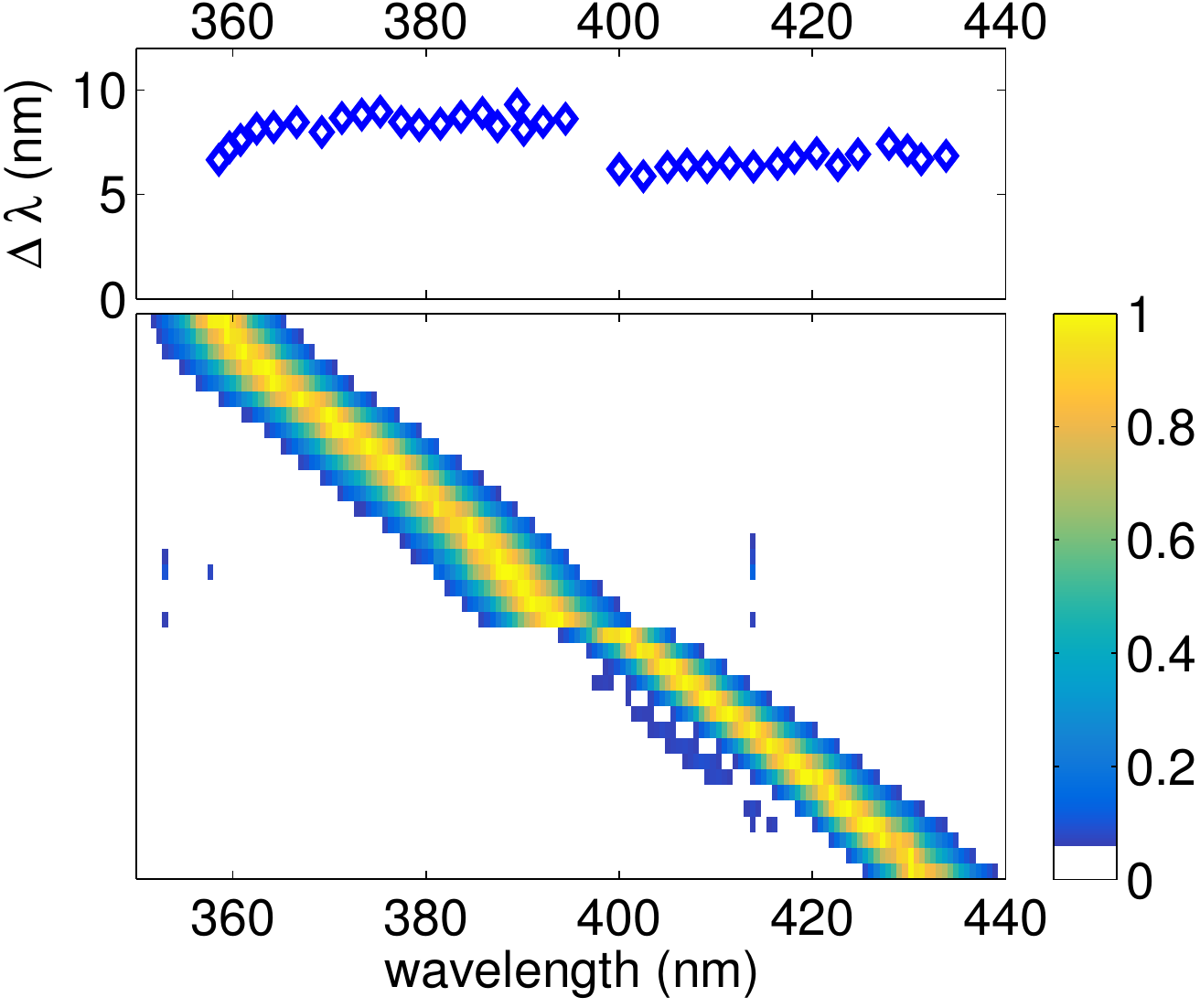}\label{fig:laserpulse_a}}
\subfigure[]
{\includegraphics[width=0.4\linewidth]{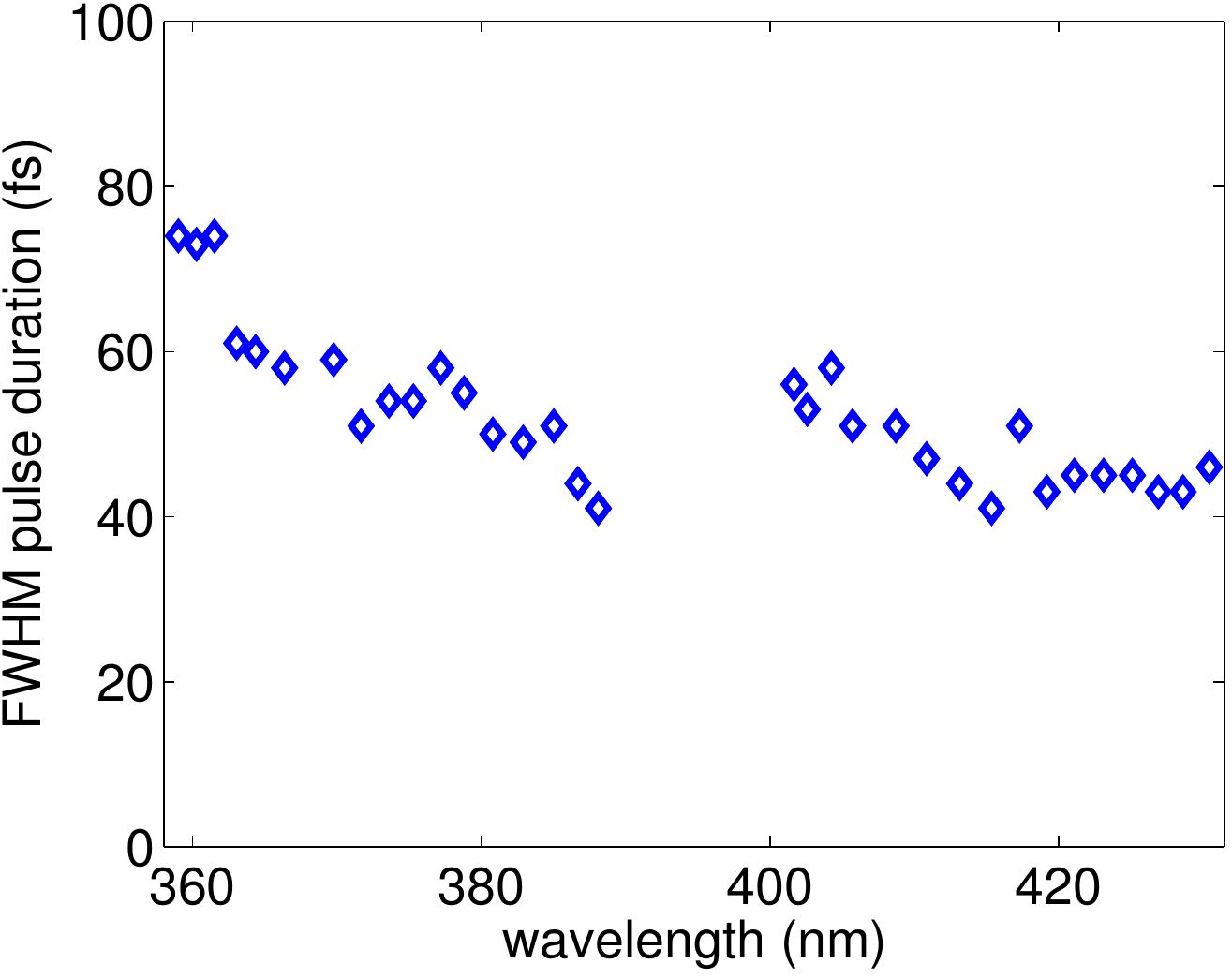} \label{fig:laserpulse_b}}
\subfigure[]{\includegraphics[width=0.4\linewidth]{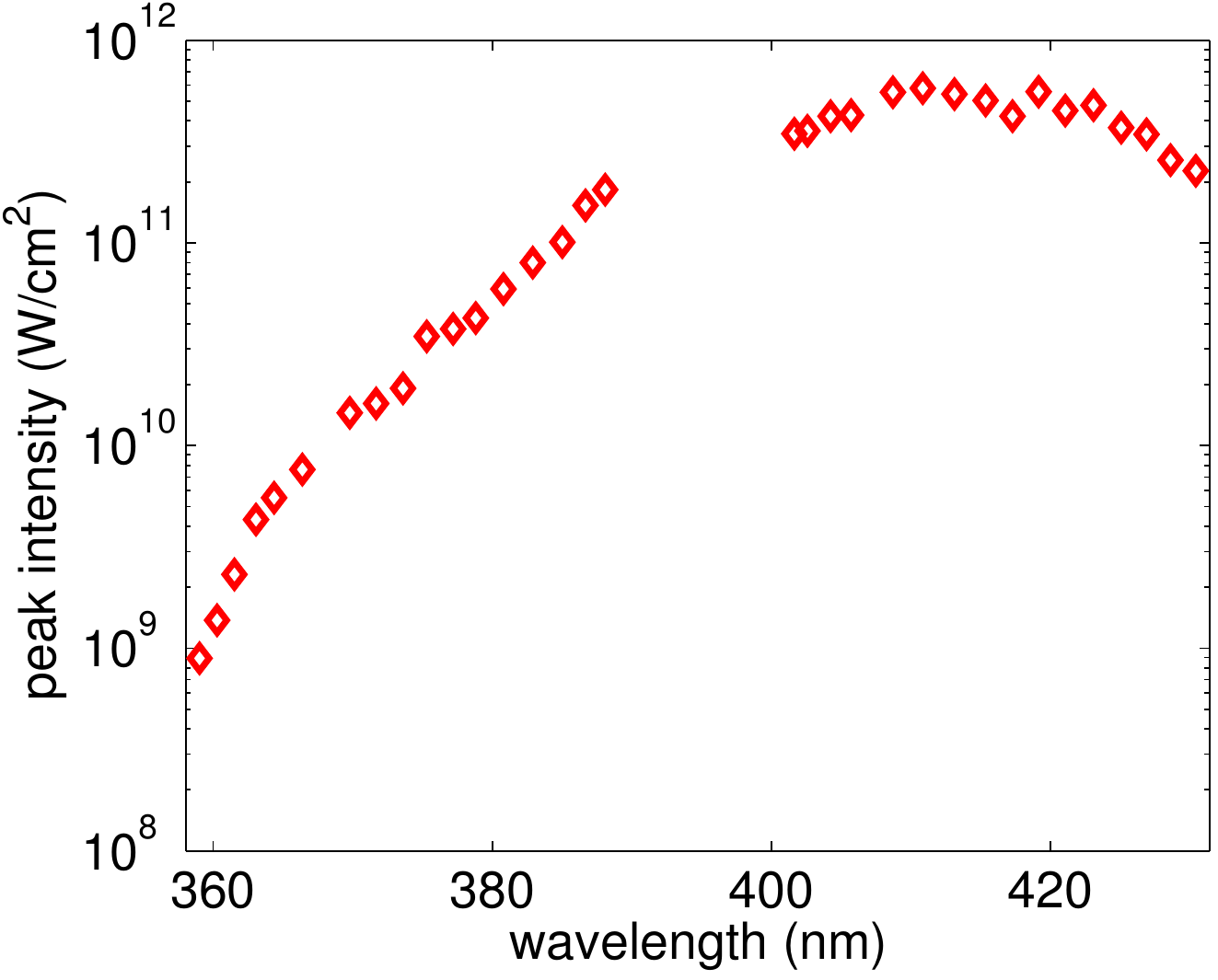}
\label{fig:laserpulse_c}}	
\subfigure[]{\includegraphics[width=0.4\linewidth]{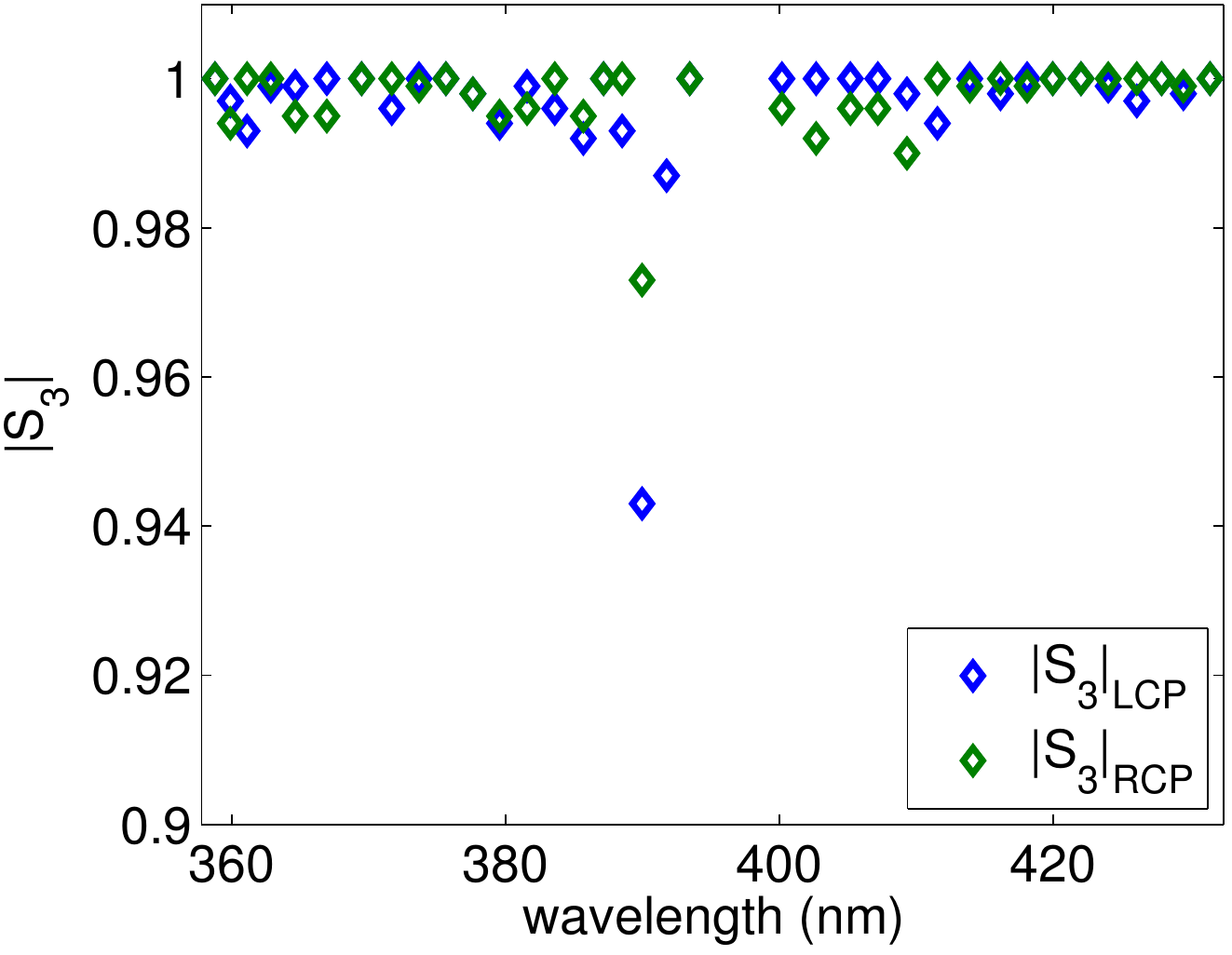}
\label{fig:laserpulse_d}} \\
		\caption{Laser pulse parameter characterization. (a) Laser spectra as measured directly in front of the VMI chamber with an intensity-calibrated spectrometer (\textit{Avantes AvaSpec}). Each row represents a single measurement normalized to its maximum signal, and  the background below a signal level of 6 \% is set to white in the histogram. The discontinuity between 393 and \mbox{400 nm} originates from different optical settings inside the \textit{TOPAS}. Corresponding FWHM bandwidths are shown above. (b) FWHM pulse durations measured by TG FROG for different excitation wavelengths. (c) Using the recorded laser power and focused beam diameter (measured by a \textit{WinCamD}), the peak intensity is estimated and plotted as a function of wavelength (shown on a logarithmic scale). (d) Characterization of the Stokes |S$_3$| parameter.  }		\label{fig:laserpulse}
\end{figure}

The schematic experimental setup of the femtosecond PECD experiment is depicted in figure \ref{fig:ExpSetup}. A titanium sapphire amplifier (\textit{Femtopower HE \mbox{5 kHz}}) generating laser pulses with 5 kHz repetition rate, a center wavelength of 785 nm, a duration of about \mbox{25 fs} and 1 mJ pulse energy is used to drive frequency conversion in a tunable collinear optical parametric amplifier (\textit{TOPAS, Light Conversion}). The laser wavelength can be tuned from the UV to the near-infrared spectral region. Compensation for accumulated dispersion of the UV laser pulses is achieved by a prism compressor providing short pulses in the interaction region. The wavelength scan was performed in thirty-six steps in the region between 359 and \mbox{431 nm}, amounting to a mean step size of about \mbox{2 nm} which is small compared to the bandwidth of the laser pulses (depicted at the top of figure \ref{fig:laserpulse_a}). The laser spectra as recorded directly in front of the VMI chamber using an intensity-calibrated spectrometer (\textit{Avantes AvaSpec}) are depicted in figure \ref{fig:laserpulse_a}.

The high-frequency cutoff of the scanning range is limited by the laser power, which decreases significantly below 359 nm (see figure \ref{fig:laserpulse_c}). The low-frequency cutoff at 431 nm lies near the threshold for three-photon ionization out of the HOMO of fenchone. \\ \indent
The center wavelength of each spectrum is found by fitting a Gaussian to the data and is taken as experimental wavelength for the respective data in the further evaluation. We typically obtain frequency-converted laser pulses of about 50 fs duration and 6\textendash 9 nm bandwidth (see figure \ref{fig:laserpulse_b}). The pulse duration at each wavelength was measured independently with a home-built transient grating frequency resolved optical gating \cite{Trebino.1997} (TG FROG) and retrieved by Trebino's \textit{MATLAB} algorithm. The center wavelength assigned to each FROG measurement was determined by measuring the laser spectrum in front of the FROG with the same intensity-calibrated spectrometer as in the VMI experiment and fitting a Gaussian to the data. The beam path for the FROG measurements was similar to the one used in the VMI experiment although the focusing lens was not included and the vacuum chamber window (thickness 5 mm) was modeled by combination of a 4 mm plate of the same material and the beam path in air required for the FROG. The largest relative deviation in wavelength between the FROG and the VMI experiment is about 0.4 \%, much less than the smallest bandwidth of the laser. Three points of the FROG measurements in the region between 390 and 400 nm showed a deviation in wavelength comparable to the laser bandwidth and are therefore not shown, although the pulse durations measured were similar to adjacent points. The prism compressor was optimized using the FROG trace for minimum pulse duration for each wavelength setting to estimate minimum pulse duration reachable in the VMI experiment. During the VMI experiment, the prism compressor was optimized using the total photoelectron yield. \\ \indent
The laser power was measured by an \textit{Ophir Nova II} and the focused beam diameter  was measured by a \textit{WinCamD}. The measured pulse duration, laser power, and focused beam diameter were used to estimate the focal intensities (see figure \ref{fig:laserpulse_c}). Note that intensities at 398 nm used in a previous scan spanning $(2.0 \times 10^{12}$ \textendash $1.1 \times 10^{13}) \frac{\text{W}}{\text{cm}^2}$, \cite{Lux.2015b} showed only minor influence on PECD. The effect was different but still observable in the tunneling regime. \cite{Beaulieu.2016a} The intensities used in this work are far below $10^{13} \frac{\text{W}}{\text{cm}^2}$, so we do not expect strong influence of intensity on PECD. \\ \indent
Horizontal polarization in front of the VMI chamber was assured via polarization measurements. An achromatic quarter-wave plate (QWP, \textit{B.Halle}) was used to convert linearly polarized (LIN) to left circularly polarized (LCP) or right circularly polarized light (RCP). The degree of polarization is determined using a Glan-Laser polarizer (\textit{ThorLabs}) and a powermeter (\textit{Ophir Nova II}). The quality of circular polarization is quantified by the Stokes |S$_3$| parameter and is\textemdash with the exception of one measurement point, where the laser power was unstable\textemdash \ well above 98 \% throughout the scanning range (see figure \ref{fig:laserpulse_d}). \\ \indent
The laser beam is focused into the interaction region of a VMI spectrometer by an \mbox{$f = 200 $ mm} plano-convex lens and intersected with an effusive gas beam of fenchone. The three-dimensional photoelectron momentum distribution is projected by a VMI setup consisting of three stainless steel plates onto an imaging detector comprising an MCP assembly in Chevron configuration and a phosphor screen (\textit{SI-Instruments GmbH}). The projected photoelectron angular distributions (PADs) are recorded using a 1.4 million pixel CCD camera (about 12 bit, \textit{Lumenera Lw165m}). Alternatively, ion TOF mass spectra can be recorded on an oscilloscope (\textit{LeCroy Waverunner 640Zi}) via a capacitively coupled output. \\ \indent
For each wavelength setting of the \textit{TOPAS}, we average a total of 2500 images of the camera each for LIN, LCP, and RCP. During the RCP and LCP measurements, the polarization was reversed every 500 images of the camera to reduce the effects of slow experimental fluctuations. The total measurement time is about one minute ($\sim$ 1.5 million laser pulses). The PECD image is calculated by subtracting the RCP PAD image from the LCP PAD image. \\ \indent
To quantify contributions from different ionization channels, the original three-dimensional photoelectron distribution is reconstructed via an Abel inversion. We use the pBasex algorithm, \cite{Lux.2015b} expanding the PAD images into a series of Legendre polynomials truncated after the 8$^{\text{th}}$ order following Yang's \mbox{theorem. \cite{Yang.1948}} The pBasex algorithm yields Legendre coefficients for all photoelectron energies visible on the detector and thereby contributions from different ionization channels can be separately evaluated. The PECD magnitude can be quantified via a sum over the contributions of the odd-order Legendre coefficients ($c_i$) divided by the total signal ($c_0$), denoted as linear PECD (LPECD): \cite{Lux.2015b} 

\begin{equation}
\text{LPECD} = \frac{1}{c_0} \left(2 c_1 - \frac{1}{2} c_3 + \frac{1}{4} c_5 - \frac{5}{32} c_7\right).
\label{equ:LPECD}
\end{equation}

For the case of non-alternating signs between the odd-order coefficients, the quadratic PECD (QPECD) proves more helpful for quantification: \cite{Lux.2015b}

\begin{equation}
\text{QPECD} \approx \frac{\sqrt{12}}{c_0} \sqrt{\left( \frac{1}{3}c_1^2 + \frac{1}{7} c_3^2 + \frac{1}{11} c_5^2 + \frac{1}{15} c_7^2 \right)} .
\label{equ:QPECD}
\end{equation}

\section{Results and Discussion}

\subsection{Investigation of intermediate resonances via nanosecond 2+1 REMPI}
\label{sec:nsREMPI}

\begin{figure}
\includegraphics[width=0.9\linewidth]{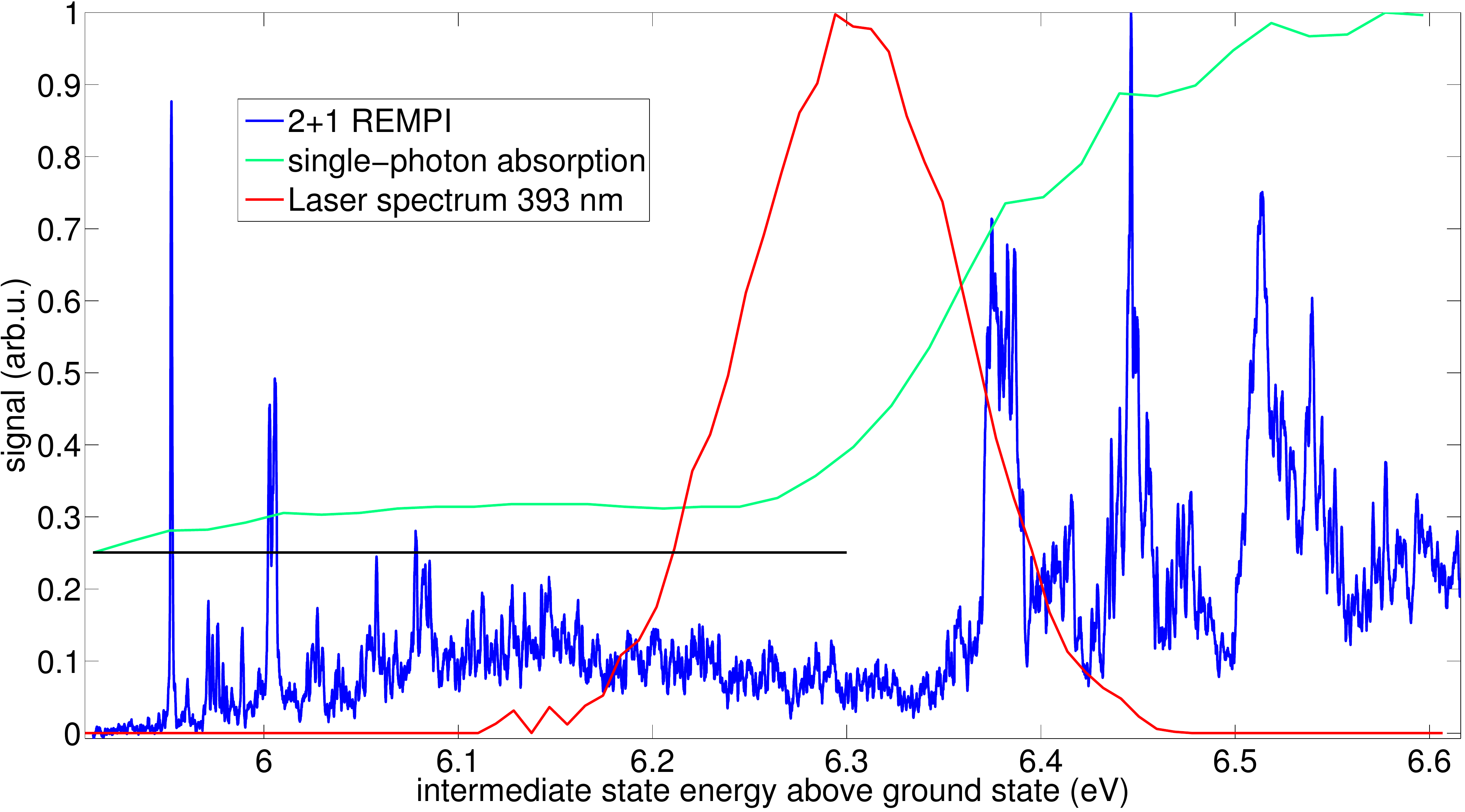}
 \caption{Comparison between absorption measurements at room temperature \cite{Pulm.1997} (green) and high-resolution 2+1 REMPI spectra of (S)-(+)-fenchone from a cold molecular beam (blue). The absorption curve is shifted upwards for better visibility and the black line beneath marks its baseline. Previous 2+1 REMPI measurements \cite{Driscoll.1991} covering the 5.9\textendash 6.2 eV range are reproduced by the blue curve, and we extended the energy range so that contributions from the adjacent resonance can also be investigated. For comparison, a frequency-doubled pulse profile of the femtosecond laser with a center wavelength of 393 nm is shown in red.  This is the point at which the second intermediate resonance appears in the femtosecond experiment. }
	\label{fig:Absspectra}
\end{figure}

Single-photon absorption measurements on bicyclic ketones at room temperature have been carried out in earlier work, \cite{Pulm.1997} providing an estimate of the energies of Rydberg states in the 3.5\textendash 8.5 eV range. These measurements can be compared to previous 2+1 REMPI studies on the same group of molecules  \cite{Driscoll.1991} in which the fenchone spectrum was reported in the 5.9\textendash 6.2 eV range. As this energy range covers only the first intermediate resonance sampled in our femtosecond wavelength scan (the 3s state), we repeated this experiment for a broader wavelength range at the University of Göttingen (see figure \ref{fig:Absspectra}) using the setup described in section \ref{sec:nsExpSetup}. \\ \indent
Due to the narrow-bandwidth excitation and rovibrationally cold molecular beam source, rather sharp features were found. If we compare the synchrotron absorption measurements in figure \ref{fig:Absspectra} (green) with our high-resolution 2+1 REMPI measurements (blue) we notice that the multi-photon excitation qualitatively follows the single-photon absorption but is able to resolve the fine structures of intermediate resonances. 
At 6.37 eV, a sharp rise in intensity is observed, which is accompanied by an apparent increase in linewidth. This suggests that the feature at 6.37 eV might reasonably be assigned to the 0-0 transition of the C-band (3p $\leftarrow$ n). The peak at \mbox{6.37 eV} is reproduced in the absorption spectrum. \cite{Pulm.1997} Our measured difference between the origin of the B-band (3s $\leftarrow$ n, 5.95 eV) and the C-band (3p $\leftarrow$ n, \mbox{6.37 eV}) is \mbox{0.42 eV}, which is in reasonable agreement with the difference between the maxima of the B- and C-band envelopes (\mbox{0.48 eV}) obtained from absorption studies (plotted in green in \mbox{figure \ref{fig:Absspectra}). \cite{Pulm.1997}}  \\ \indent
We therefore assign the peaks starting at 5.95 eV to the 3s and the ones starting at \mbox{6.37 eV} mainly to the 3p state. The 3s and 3p state can be populated simultaneously in the femtosecond experiment e.g. by using the laser pulse centered at \mbox{393 nm}, whose spectral profile is plotted in figure \ref{fig:Absspectra} (red). Within each of the two states three prominent peaks are seen in the 2+1 spectrum spaced by about 500 cm$^{-1}$. \\ \indent
Evaluation of the first pronounced peak of the 3s state by fitting a Lorentzian \cite{Demtroder.2011} to the data yields a lifetime of about 1.3 ps (1/e) matching coarsely the observed 3.3 ps decay time for the total photoelectron signal. \cite{Comby.2016}  
In that work this ps lifetime was attributed to internal conversion from the 3s state to the ground state. In addition a 400 fs time scale was found on the first odd Legendre coefficient and suggested to be an indication for  vibrational relaxation dynamics approaching the equilibrium geometry of the
3s state. Note within that context, that in our high-resolution data there is an unstructured background which is consistent with femtosecond time scale energy redistribution dynamics. \cite{Baumert.1993} Also note the increase in width of the observed peaks in the 3p as compared to the 3s state, which might suggest about one order of magnitude shorter lifetime for the 3p state.

\subsection{Femtosecond PECD}
\label{sec:fsPECD}

\subsubsection*{Energy scaling and intermediate states}

The enantiopure fenchone samples were purchased from Sigma-Aldrich with a specified purity of \mbox{99.2 \%} and used without further purification. The enantiomeric excess (e.e.) as measured by gas chromatography was \mbox{99.9 \%} for (S)-($+$)-fenchone and 84 \% for \mbox{(R)-($-$)-fenchone}. \cite{Kastner.2016} Due to the higher purity, the wavelength scan was performed on (S)-($+$)-fenchone. We checked the mirroring of forward-backward asymmetry when switching to (R)-($-$)-fenchone at four wavelength settings and observed the expected behavior.

\begin{figure}
\centering
\includegraphics[width=\linewidth]{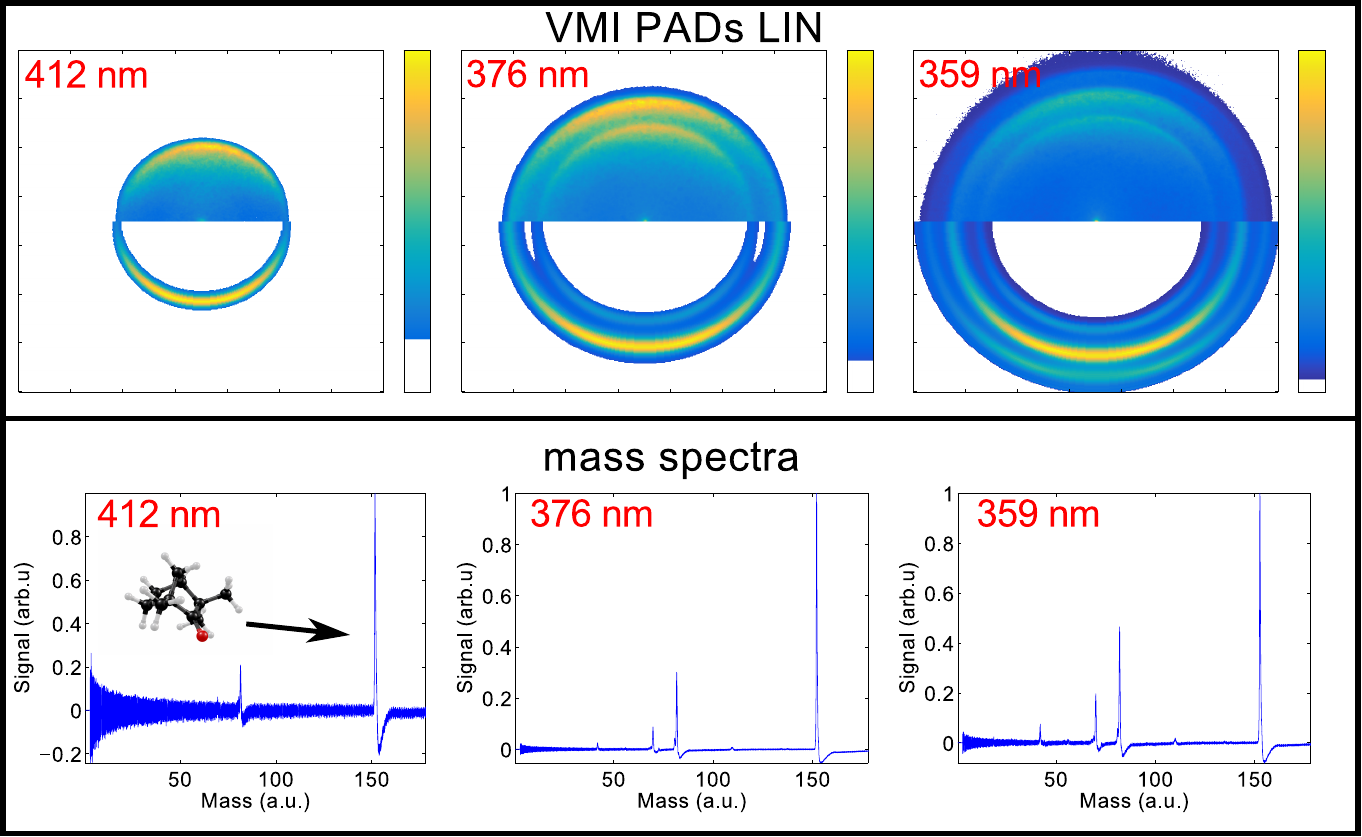}
\caption{
\label{fig:PADsionTOF}
Top row: VMI PAD raw images for LIN (upper half) and Abel inverted cuts (lower half of image) obtained at excitation  wavelengths of 412, 376 and 359 nm. The different contributions in the photoelectron spectrum arise from excitation of the B-band (3s $\leftarrow$ n) at 412 nm, B- and C-band \mbox{(3p $\leftarrow$ n)} at 376 nm and B-, C- and $\pi^* \leftarrow \sigma $ excitation at \mbox{359 nm}. The detailed discussion of the assignment of intermediates can be found in the text. Bottom row: Corresponding mass spectra. }
\end{figure}

\begin{figure}
\centering
    \subfigure[]{\includegraphics[width=0.45\linewidth]{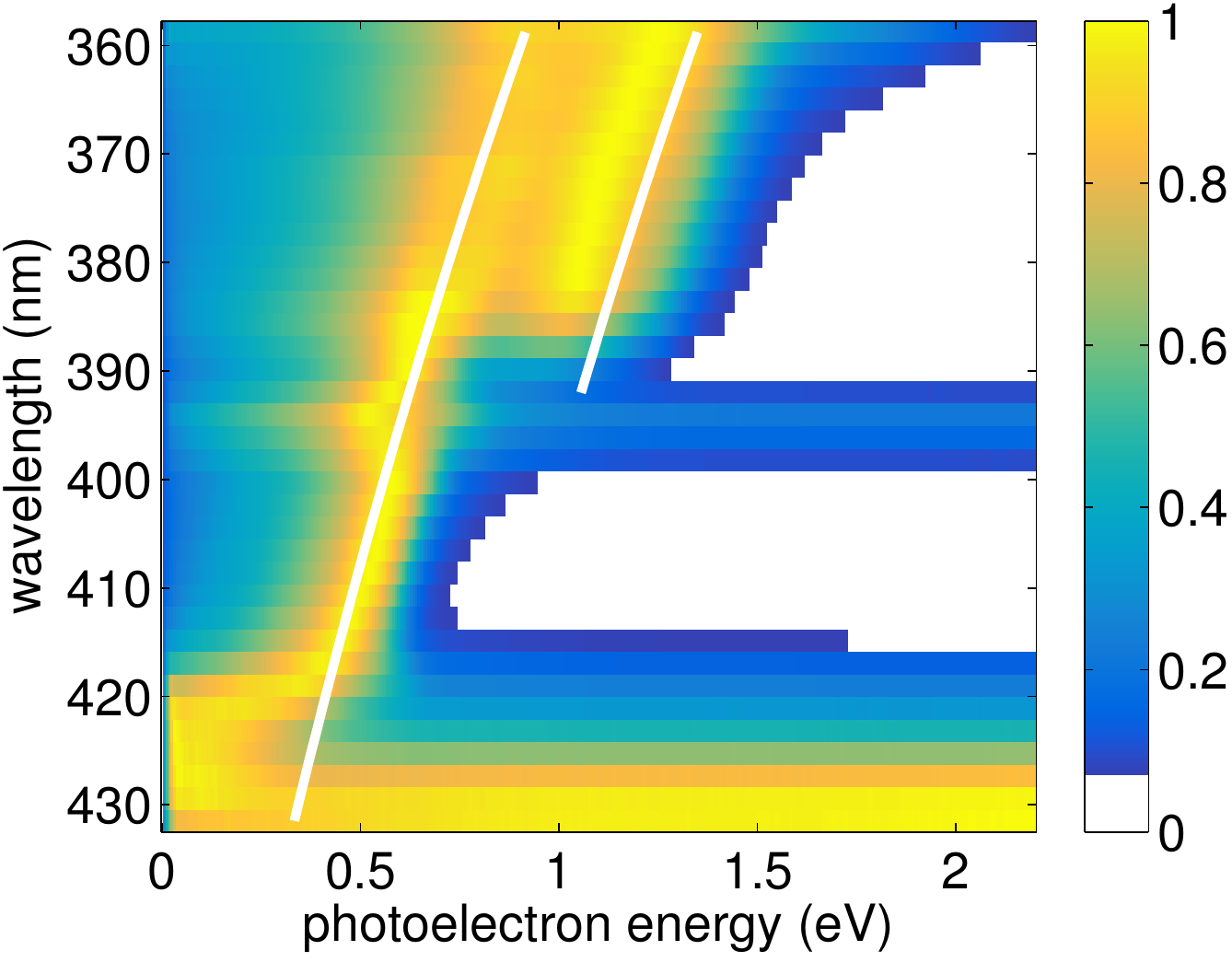}\label{fig:Energyplot_LIN_a}}
\hfill
\subfigure[]{	\includegraphics[width=0.45\linewidth]{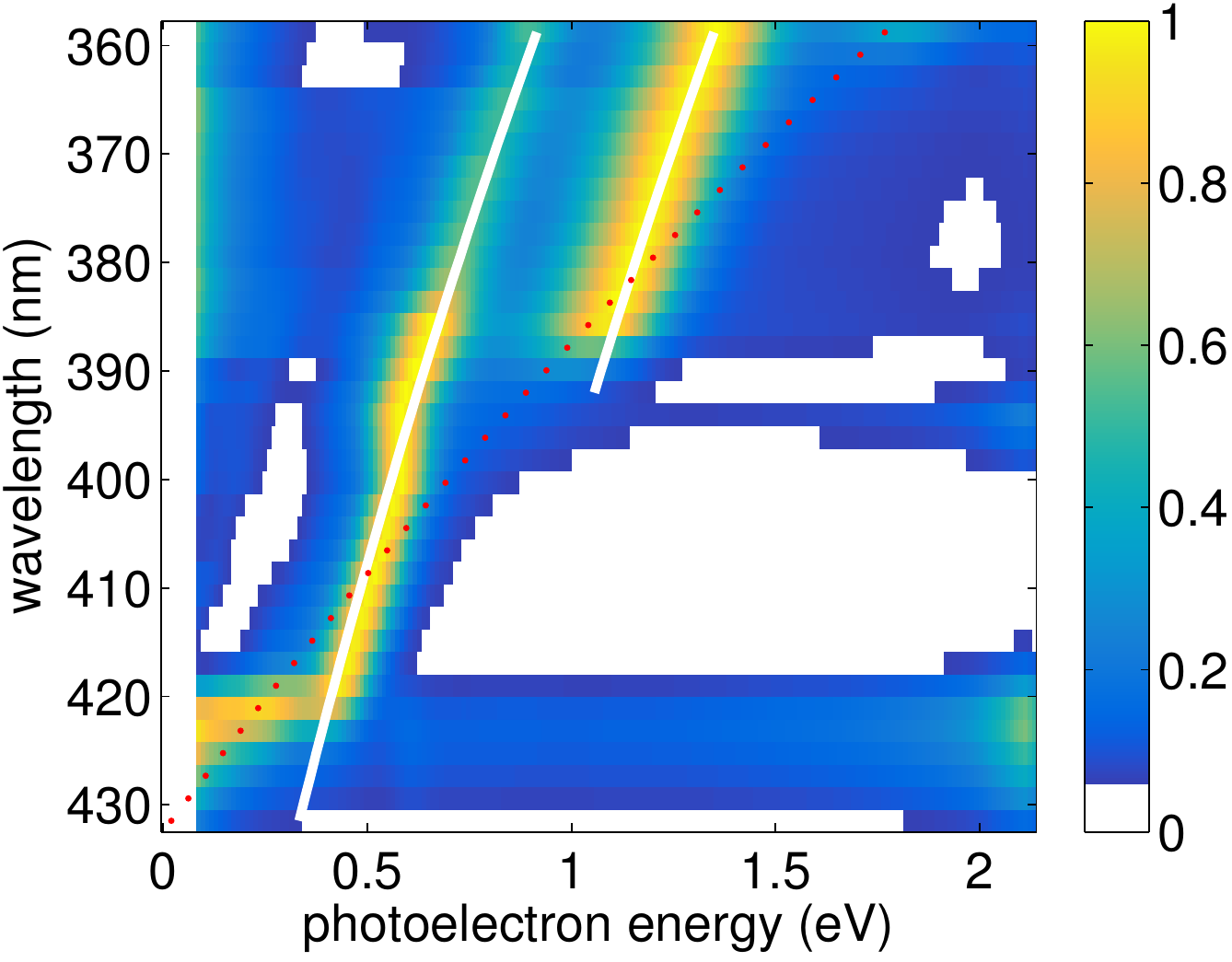}	\label{fig:Energyplot_LIN_b}}			\caption{(a) Photoelectron spectra derived from the PAD raw images and (b) from the cuts through the reconstructed three-dimensional photoelectron distributions by the pBasex for LIN of (S)-($+$)-fenchone. Every row represents a single measurement. The white lines in the plots indicate the best fit to the Abel inverted photoelectron spectra showing an energy scaling with $\hbar \omega$. Owing to the Abel projection, the peaks for the photoelectron spectra derived from the PAD raw images in (a) are shifted towards lower energies. Additionally, the expected behavior for a non-resonant three-photon energy scaling is indicated by the dotted red line in (b). }		\label{fig:Energyplot_LIN}
\end{figure}

We performed intensity scans on (R)-($-$)-fenchone at an excitation wavelength of \mbox{389 nm} yielding a power law between 2.5 (fenchone parent ion TOF peak) and 3.1 (total photoelectron yield) supporting the assumption of a dominant three-photon process. \cite{Lux.2015b} Furthermore, a ponderomotive shift of photoelectron energy with laser intensity was not observed indicating that the intermediate states act as Freeman resonances as described in previous work. \cite{Lux.2015b} \\ \indent
The VMI voltages in the measurements were set to image photoelectrons having energies up to about \mbox{4.3 eV} onto the detector. Synchrotron experiments on fenchone in the valence region \cite{Powis.2008b} provide the position of the HOMO at 8.6 eV and the HOMO$-$1 at 10.4 eV  below the ionization energy. Three-photon excitation by 359 nm corresponds to an energy of 10.36 eV, which is slightly below the threshold for ionization out of the HOMO-1. In our scanning range, we observe no photoelectrons or PECD corresponding to four-photon ionization out of the HOMO$-$1. The observed contributions in the photoelectron spectra can thereby be explained by three-photon ionization out of the HOMO. The intensities used herein seem to be insufficient to give rise to four-photon ionization of fenchone. This is consistent with the fact that our experiments used less intensity than recent experiments probing four-photon HOMO$-$1 ionization or above-threshold ionization. \cite{Beaulieu.2016a, Beaulieu.2016b, Lux.2016}  The PAD images are therefore cropped for the Abel inversion to an image size that covers energies up to about 2.1 eV.  \\ \indent
Previous experiments have investigated ionization via the 3s Rydberg state. \cite{Lux.2015b, Beaulieu.2016b} In the current experiment, we also observe contributions from intermediate states lying higher in energy than the 3s in the photoelectron spectra as shown for excitation at 376 and 359 nm in \mbox{figure \ref{fig:PADsionTOF}}. The photoelectrons are ionized to different energies in the continuum. \\ \indent
We measured mass spectra at each excitation wavelength. Throughout the scanning range, the fenchone parent ion was by far the predominant mass signal and no significant intensity changes of the fragment ions were found. We therefore conclude that the different lifetimes of intermediate states derived from high-resolution 2+1 REMPI (see section \ref{sec:nsREMPI}) have only minor influence on fragmentation of the molecules. Representative mass spectra for excitation at 412, 376 and 359 nm are shown in the bottom row of figure \ref{fig:PADsionTOF}. \\ \indent
To determine the dependence of photoelectron energy on excitation wavelength, we consider the Abel inverted photoelectron spectra in figure \ref{fig:Energyplot_LIN_b} derived from the PAD raw images for LIN in figure \ref{fig:Energyplot_LIN_a}. The PAD raw images and the cuts through the Abel inverted PADs are transformed to polar representation. We applied signal normalization by corresponding area to the photoelectron spectra derived from the Abel inverted cuts, and contributions below \mbox{80 meV} were set to white in the histogram as the normalization factor grows with decreasing radius. Energy calibration is applied as described before \cite{Lux.2015b} and the PADs are corrected in signal for the transformation from momentum to energy representation. To improve dynamic range in both images, each row is normalized to its maximum and weak signals below a level of 6 \% are set to white in the histogram. For wavelengths below \mbox{422 nm} in the Abel inverted cuts the energy region for finding the maximum is restricted to values above \mbox{80 meV}. \\ \indent
In the spectral region between 431 and 422 nm\textemdash near the threshold for three-photon ionization\textemdash \ the distribution in the photoelectron spectra is rather broad, the overall signal is very weak and the broad feature can be attributed to accumulating noise. Between 422 and \mbox{393 nm}, a single energy ring can be seen which originates from ionization via the 3s state. Below 393 nm an additional energy ring originating from ionization via the 3p state appears and becomes more pronounced with decreasing wavelength. A third contribution arises below 363 nm originating from ionization via the $\pi^* \leftarrow \sigma $ transition (see also \mbox{figure \ref{fig:PADsionTOF}}). The cuts through the three-dimensional photoelectron distributions (depicted in figure \ref{fig:Energyplot_LIN_b}) using the pBasex algorithm provide information on the dependence of photoelectron energy on wavelength. The local maxima of the reconstructed photoelectron spectra for the three contributions are investigated in the spectral region between 359 and 422 nm for the first, between 359 and \mbox{389 nm} for the second and  between 359 and 363 nm for the third contribution.

The observed scaling of photoelectron energy with wavelength implies that the third photon determines the photoelectron energy in agreement with recent observations on fenchone \cite{Beaulieu.2016b} and similarly on camphor \cite{Lehmann.2013} as well as on limonene. \cite{Fanood.2016} This corroborates the assumption that the two first contributions in the photoelectron spectra arise from excitation of different Rydberg states, where the potential energy surfaces of intermediates and cation are quasi-parallel. The ionization step is thereby governed by $\Delta v = 0$ transitions. Thus, the photoelectron energy scales as $\hbar \omega$, where $\omega$ is the angular laser frequency. The high-resolution 2+1 REMPI experiment demonstrated that both Rydberg states exhibit long Franck-Condon progressions. \\ \indent
Taking the linear scaling into account, the local maxima in the Abel inverted photoelectron spectra assigned to the $i$-th intermediate state ($i=1,2,3$) are fitted by an expression $\frac{hc}{\lambda}-(IP^{Fen}_{\perp}-E_{int,i})$, where $\lambda$ denotes the laser wavelength, $IP^{Fen}_{\perp}$ is the vertical ionization potential (IP) of fenchone and $E_{int,i}$ denotes the energy of the intermediate state $i$. The best fits for the 3s and 3p state are depicted as white lines in figure \ref{fig:Energyplot_LIN}. As the energy needed for vibrational excitation is transferred to the cation it does not manifest in the photoelectron spectrum so that $E_{int,i}$ for i=1,2 denotes the energy of the ground vibrational level of the respective Rydberg state. \\ \indent
Using $IP^{Fen}_{\perp} = $ 8.6 eV \cite{Powis.2008b} and the standard deviation of the data as error metric, the energies of intermediate resonances are found to be $(6.059 \pm  0.017)$ eV, \mbox{$(6.493 \pm 0.015)$ eV} and $(6.935 \pm 0.019)$ eV for the first, second and third resonance, respectively. The energy separation between the first two resonances in the femtosecond excitation of about 0.43 eV matches the difference found in high-resolution 2+1 REMPI of about 0.42 eV well. As the energy separations from the ground state are in quite good agreement with previous experimental \mbox{values \cite{Pulm.1997}} of 6.10 and 6.58 eV, we conclude that we excite the B- (3s $\leftarrow$ n) and C-band (3p $\leftarrow$ n) during ionization. 
The derived absolute energy of intermediates is influenced by $IP^{Fen}_{\perp}$ which we would need to reduce by about \mbox{110 meV} to reach a similar value as in the high-resolution 2+1 REMPI (see section \ref{sec:nsREMPI}) for the 3s state. \\ \indent
The third contribution is only observed for the lowest four wavelength settings, but the energy of \mbox{6.94 eV} coarsely matches the expected value from DFT simulation \cite{Pulm.1997} for the lowest $\pi^* \leftarrow\sigma$ excitation at \mbox{7.05 eV}. Due to the limited number of data points, we did not include the third resonance in the PECD evaluation in the next section.

\subsubsection*{Energy dependence of PECD}

\begin{figure}
\centering
\includegraphics[width=\linewidth]{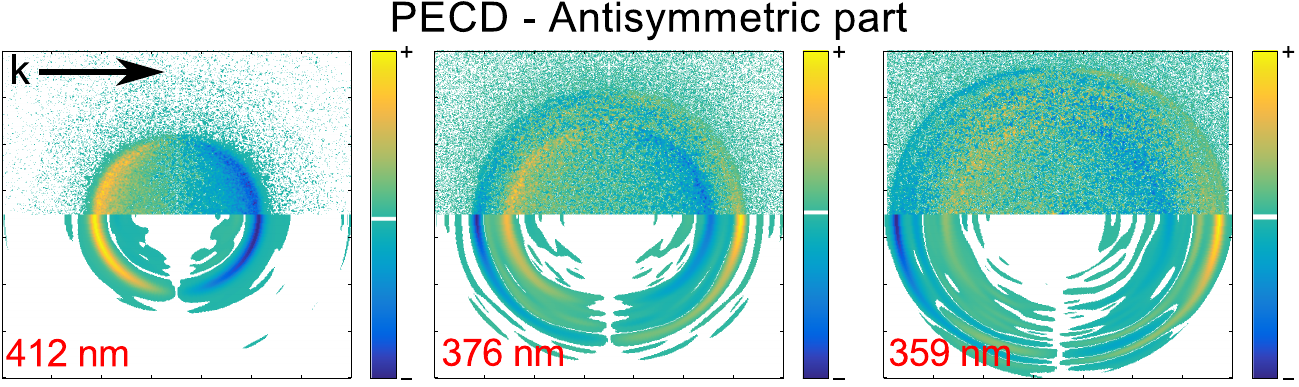}
\caption{
\label{fig:PECD_Antisymm}
The antisymmetric part of the PECD obtained at excitation wavelengths of 412, 376, and 359 nm. The top half of each image shows the raw PECD and the bottom half shows a cut through the corresponding inverse Abel transform. The different contributions in the photoelectron spectrum arise from excitation of the B-band (3s $\leftarrow$ n) at 412 nm, B- and C-band \mbox{(3p $\leftarrow$ n)} at \mbox{376 nm} and B-, C- and $\pi^* \leftarrow \sigma$ excitation at 359 nm. The inversion of forward-backward asymmetry for the contributions from 3s and 3p state is reflected in the sign change of PECD (see figure \ref{fig:PECD_b}). }
\end{figure}

PECD is a sensitive probe of the molecular potential. The observed forward-backward asymmetry depends on photoelectron energy as predicted by theory for ionization out of the valence shell. \cite{Powis.2000} Qualitatively, forward-backward asymmetry for fenchone increases for decreasing photoelectron energy when approaching the ionization threshold for single-photon ionization out of the HOMO. \cite{Nahon.2016} The single-photon experiment \cite{Nahon.2016} has been able to measure slow electron PECD down to an energy of about 0.6 eV. We are able to follow the PECD curve about 200 meV further down to a photoelectron energy of about \mbox{0.4 eV} and up to about 1.5 eV. \\ \indent
Examples of raw and Abel inverted antisymmetric parts of the PECD images \cite{Lux.2015b} are depicted in figure \ref{fig:PECD_Antisymm}.  The Legendre coefficients for the 3s state excitation are found by averaging over the FWHM width of $c_0$. For the 3p state, the averaging window is shifted upwards by \mbox{94 meV} relative to the respective $c_0$ maximum, because the PECD peaks there. A similar upward shift was observed for above-threshold ionization out of the HOMO via the 3s state. \cite{Beaulieu.2016b}

\begin{figure}
   \subfigure[]{	\includegraphics[width=0.45\linewidth]{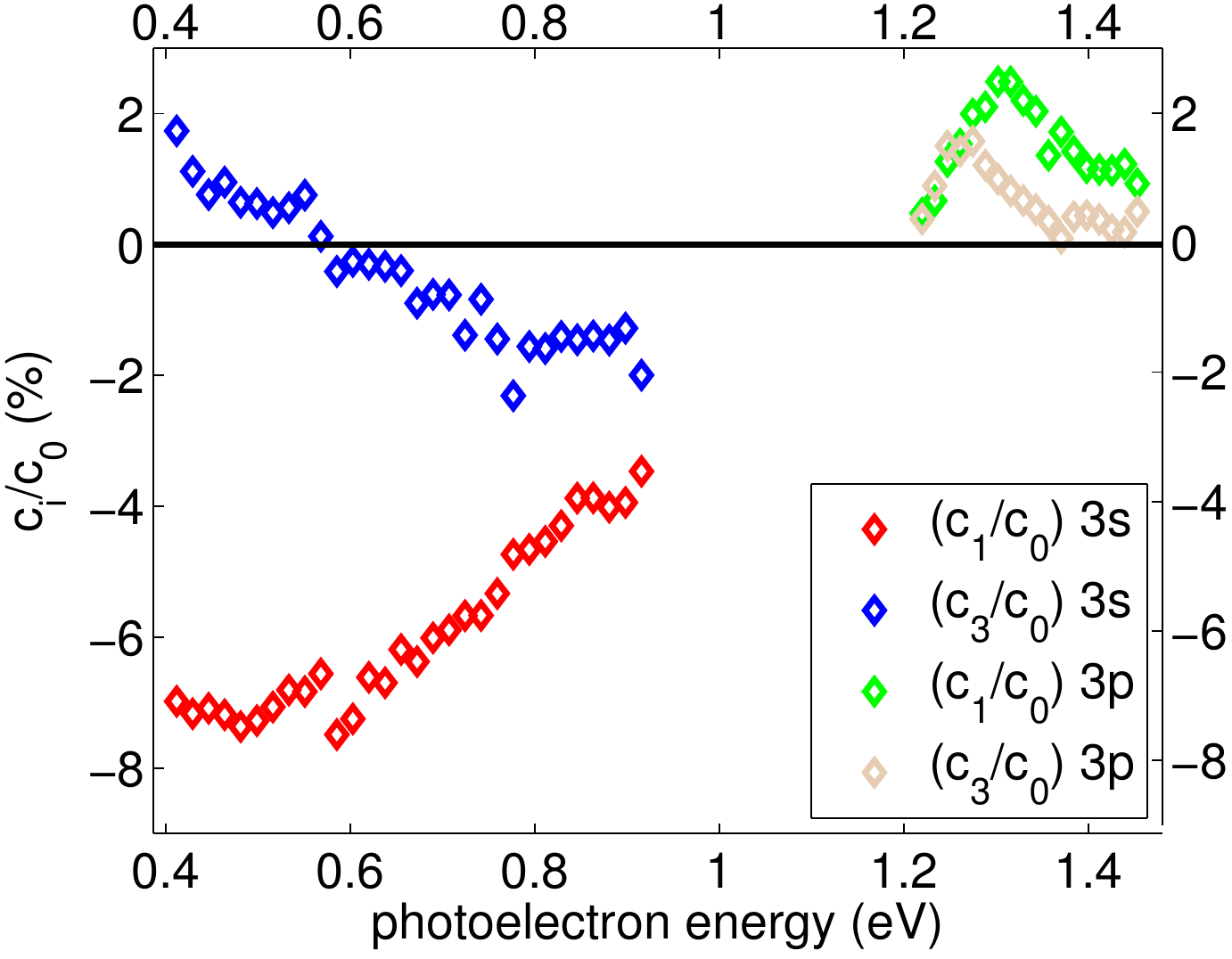}\label{fig:PECD_a}} 
   \hfill   
   \subfigure[]{	\includegraphics[width=0.45\linewidth]{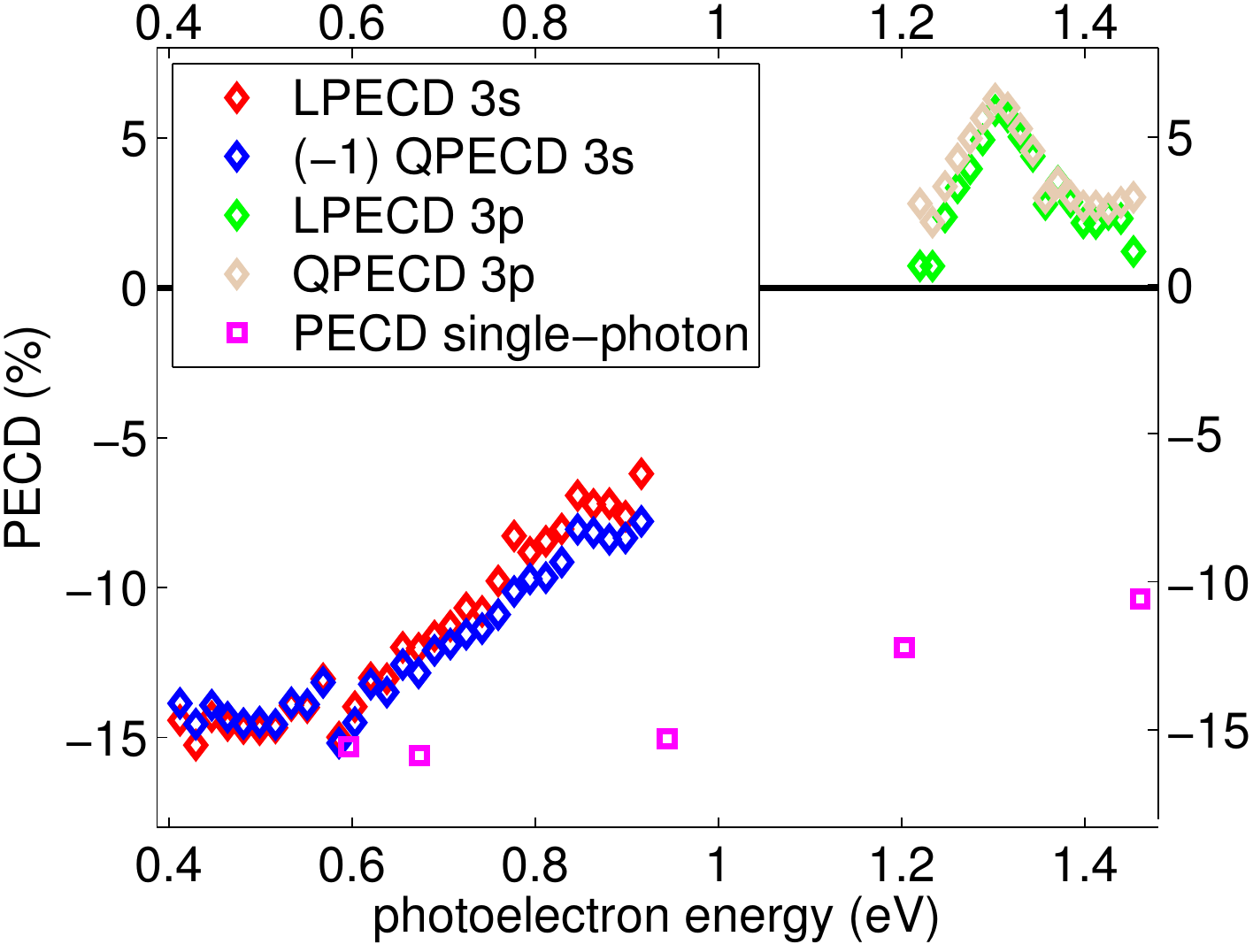}
 \label{fig:PECD_b}}				\caption{(a) $c_1$ and $c_3$ coefficients normalized to the total signal ($c_0$) and (b) corresponding PECD quantified by different metrics (see legend). Photoelectrons originating from ionization via the  3p state have higher energies as expected from the larger energy separation from the electronic ground state. For comparison, values for slow electron single-photon PECD \cite{Nahon.2016} are plotted as purple squares showing a different dependence on energy. }
\label{fig:PECD}
\end{figure}

In figure \ref{fig:PECD_a}, the averaged coefficients are plotted as a function of photoelectron energy.
$c_1$ dominates the forward-backward asymmetry throughout the energy range accessible to the photoelectrons coming from the 3s state. 
At low photoelectron energies, the $c_1$ and $c_3$ coefficients obtained from the 3s state have opposing signs. However, at a photoelectron energy of 0.56 eV, the value of the $c_3$ coefficient crosses through zero, and above this photoelectron energy both coefficients have the same sign. This leads to cancellation effects in linear PECD metrics such as LPECD. The higher order contributions $c_5$ and $c_7$ oscillate around zero and are therefore not discussed in more detail although they are included in the calculation of LPECD (see equation \ref{equ:LPECD}) and QPECD (see equation \ref{equ:QPECD}).

PECD is plotted as a function of photoelectron energy in figure \ref{fig:PECD_b} and compared to single-photon PECD results for ionization out of the HOMO of fenchone. \cite{Nahon.2016} Throughout the scanning range, LPECD and QPECD show very similar behavior. In the region close to ionization threshold we observe a similar behavior as found in single-photon ionization, namely increasing PECD for slow electrons. However, the slope of single-photon PECD is different than that of the PECD for REMPI via the 3s state. By extrapolation we find the point where the PECD curve for the 3s state crosses through zero in the 1.1\textendash \mbox{1.4 eV} photoelectron energy range. This energy is much lower than in the single-photon case, where the first zero transition is found at about 3 eV. \cite{Nahon.2016} \\ \indent
Photoelectrons arising from excitation of the 3p state have higher energies. The preferred emission direction of photoelectrons experiencing ionization via the 3p state is inverted with respect to ionization via the 3s state. This behavior can be observed in the antisymmetric part of the PECD raw images (depicted in figure \ref{fig:PECD_Antisymm}) and is reflected in the sign change of LPECD plotted in figure \ref{fig:PECD_b}.
In the case of limonene, the electronic character of intermediate resonances was found to have no strong influence on PECD. \cite{Beaulieu.2016a, Fanood.2016} In contrast, our wavelength scan on fenchone shows inversion of forward-backward asymmetry depending on ionization via the 3s or the 3p state. This is an indication that the intermediate resonance may have an influence on PECD. In addition we observe a different slope of PECD on photoelectron energy when comparing
multi-photon to single-photon ionization. These observations hint towards dependence of multi-photon PECD not only on final but also on the electronic character of intermediate states. Further investigations need to be done to unambiguously distinguish the role of the intermediate from the role of the final photoelectron energy. For example, a pump-probe experiment providing a fixed photoelectron energy while changing the intermediate resonance could give further insight.

\section{Summary and Conclusion}     

In this contribution, we present a fine scan of 2+1 REMPI driven PECD with respect to photoelectron energy using the prototypical chiral molecule fenchone. We demonstrate the existence of distinct photoelectron channels arising from REMPI via the 3s and 3p electronic states. The observed energy difference between both states in femtosecond excitation of \mbox{0.43 eV} is in good agreement with the separation found in high-resolution 2+1 REMPI (\mbox{0.42 eV}) as well as with previous synchrotron measurements (0.48 eV). \cite{Pulm.1997} \\ \indent
We observe that the sign of PECD obtained from ionization via the 3s or the 3p state is opposite. We are thereby able to demonstrate that the preferred photoelectron emission direction for a well-chosen wavelength can depend on the intermediate state populated during REMPI. By extrapolation we find the point where the PECD curve for multi-photon ionization via the 3s state crosses through zero in the 1.1\textendash \mbox{1.4 eV} photoelectron energy range. This energy is lower than in the single-photon case where the first zero crossing of the PECD curve is found at about 3 eV. \cite{Nahon.2016} The difference in slope of PECD on photoelectron energy in addition hints towards dependence not only on final but also on intermediate state. Further investigations need to be done to unambiguously clarify the effect of intermediates on PECD. \\ \indent
These results will prove helpful especially to benchmark theoretical modeling of the process. \cite{Goetz.2017, Demekhin.2015, Lein.2014, Lehmann.2013} Once a deeper understanding of the role of intermediate resonances is achieved, an increased sensitivity for chiral recognition in the gas phase via multi-photon PECD can be expected by giving additional selectivity.

\subsection{Acknowledgements}

The authors thank Dr. Christian Lux for contributions to data evaluation. 

\section*{Disclosure statement}

No potential conflict of interest was reported by the authors.

\section*{Funding}

Financial support by the State Initiative for the Development of Scientific and Economic Excellence (LOEWE) in the LOEWE-Focus ELCH is gratefully acknowledged. GBP acknowledges support from the Alexander von Humboldt Foundation.


\end{document}